\documentclass[epjc3]{svjour3}
\RequirePackage[T1]{fontenc}

\usepackage{epsfig}
\usepackage{graphicx}

\journalname{Eur. Phys. J. C}

\begin{document}


\title{Constraints on nuclear parton distributions from dijet photoproduction at the LHC}

\author{V. Guzey\thanksref{e1,addr1,addr2,addr3}
\and
M. Klasen\thanksref{e2,addr4}
}
\thankstext{e1}{e-mail: guzey\_va@nrcki.pnpi.ru}
\thankstext{e2}{e-mail: michael.klasen@uni-muenster.de}
\thankstext{c}{MS-TP-19-03}
\institute{Department  of  Physics,  University  of  Jyv\"askyl\"a, P.O.
 Box 35, 40014  University  of  Jyv\"askyl\"a,  Finland\label{addr1} \and
 Helsinki Institute of Physics, P.O.  Box  64,  00014 
 University  of  Helsinki,  Finland\label{addr2} \and
National Research Center ``Kurchatov Institute'',
Petersburg Nuclear Physics Institute (PNPI), Gatchina, 188300, Russia\label{addr3} \and Institut f\"ur Theoretische Physik, Westf\"alische Wilhelms-Universit\"at M\"unster,
Wilhelm-Klemm-Stra{\ss}e 9, 48149 M\"unster, Germany\label{addr4}
}

\date{\today}

\maketitle

\begin{abstract}

Using QCD calculations of the cross section of inclusive dijet photoproduction in Pb-Pb 
ultraperipheral collisions in the LHC kinematics as pseudo-data, we 
study the effect of including these data using the Bayesian reweighting technique on nCTEQ15, nCTEQ15np, and EPPS16 
nuclear parton distribution functions (nPDFs). 
We find that, depending on the assumed error of the pseudo-data, it leads to a significant reduction of the nPDF uncertainties
at small values of the momentum fraction $x_A$. Taking the error to be 5\%, the uncertainty of 
nCTEQ15 and nCTEQ15np nPDFs reduces approximately by a factor of two at $x_A=10^{-3}$. 
At the same time, the reweighting effect on EPPS16 nPDFs is much smaller due to the higher value of the tolerance and a more flexible parametrization form.

\end{abstract}

\section{Introduction}
\label{sec:intro}

Collinear nuclear parton distribution functions (nPDFs) are fundamental quantities of 
Quantum Chromodynamics (QCD) encoding information on the one-dimensional distributions of quarks and gluons 
in nuclei in terms of the light-cone momentum fraction $x_A$ at a given resolution scale $\mu$. 
Nuclear PDFs are essential ingredients of QCD calculations of cross sections at high energies involving 
charged lepton--nucleus and neutrino--nucleus deep inelastic scattering (DIS) with fixed targets and -- in the future -- in the collider mode
and proton--nucleus and nucleus--nucleus scattering at the Relativistic Heavy Ion Collider (RHIC) and the 
Large Hadron Collider (LHC). While nPDFs are non-perturbative quantities, which cannot be calculated from first principles of QCD, 
the QCD collinear factorization for hard processes and the 
Dokshitzer--Gribov--Lipatov--Altarelli--Parisi (DGLAP) evolution equations allow one to set up a framework of 
global QCD fits, which enables one to extract nPDFs from available data~\cite{deFlorian:2003qf,Hirai:2007sx,Eskola:2009uj,deFlorian:2011fp,Kovarik:2015cma,Khanpour:2016pph,Eskola:2016oht}.
Different analyses give noticeably different predictions for nPDFs and carry significant uncertainties originating mostly from the limiting kinematic coverage of the available data, indirect determination of the gluon nPDF from the DIS data using the scaling violations, and different assumptions about the shape of nPDFs at the input scale. As a result, quark nPDFs for small $x$ and the gluon nPDFs for essentially all $x$ are rather poorly known. 

Further progress in constraining nPDFs relies on studies of high-energy hard processes with nuclei at collider energies, notably, in proton--nucleus ($pA$) scattering at the LHC~\cite{Salgado:2011wc,Paukkunen:2018kmm,Citron:2018lsq} and lepton--nucleus ($eA$) 
scattering at the future EIC~\cite{Accardi:2012qut,Aschenauer:2017oxs} and LHeC~\cite{AbelleiraFernandez:2012cc}. 
However, the QCD analyses of the data on various hard processes in $pA$ scattering at the LHC during Run 1~\cite{Eskola:2013aya,Helenius:2014qla,Armesto:2015lrg,dEnterria:2015mgr} showed that the data provide only modest restrictions on nPDFs at small $x$. 
At the same time, it was proposed~\cite{Brandt:2014vva} that measurements of low-mass lepton pair production in proton--lead 
collisions at the LHC has a large potential to reduce the theoretical uncertainties on nPDFs in a wide range of $x$
 or even rule out some parameterizations. While the potential of hard $pA$ scattering at the LHC will certainty continue to be explored,
 see, e.g.~\cite{Eskola:2018sxu}, it is topical to study complementary probes of nPDFs at the LHC.
 
It has been realized that collisions of ultrarelativistic ions at large impact parameters, when the strong interaction is suppressed and the ions interact electromagnetically via the emission of quasi-real photons in so-called ultraperipheral collisions (UPCs), give an opportunity to study photon--photon, photon--proton, and photon--nucleus scattering at unprecedentedly high energies~\cite{Baltz:2007kq}. This program was realized during Run 1 at the LHC by measuring exclusive photoproduction of charmonia ($J/\psi$ and $\psi(2S)$ vector mesons) in Pb-Pb UPCs at $\sqrt{s_{NN}} = 2.76$ TeV by the ALICE~\cite{Abbas:2013oua,Abelev:2012ba,Adam:2015sia}  and 
CMS~\cite{Khachatryan:2016qhq} collaborations, 
which probes the small-$x$ gluon distribution of the target~\cite{Ryskin:1992ui}. The analyses~\cite{Guzey:2013xba,Guzey:2013qza} 
of these data in leading-order (LO) QCD gave first direct and weakly model-dependent evidence of large nuclear gluon shadowing 
down to $x \approx 10^{-3}$, which agrees very well with the predictions of the leading twist nuclear shadowing model~\cite{Frankfurt:2011cs} and the the EPS09 \cite{Eskola:2009uj}, nCTEQ15
\cite{Kovarik:2015cma}, and EPPS16 \cite{Eskola:2016oht} nPDFs. 
Note that in next-to-leading order (NLO) perturbative QCD, corrections for this process are large~\cite{Ivanov:2004vd,Jones:2015nna}
and the relation between the gluon PDF and the generalized gluon PDF is model-dependent, which makes it challenging to include the data on $J/\psi$ photoproduction on nuclei into global QCD fits of nPDFs.

During Run 2 at the LHC, in addition to light and heavy vector meson photoproduction in UPCs~\cite{Klein:2017vua,Kryshen:2017jfz,Guzey:2016piu}, the ATLAS collaboration for the first time measured inclusive dijet photoproduction in Pb-Pb
UPCs~\cite{Atlas}.
Predictions for rates of this process in $pA$ and nucleus--nucleus ($AA$) UPCs at the LHC in LO QCD with an emphasis of heavy quark production were made in Ref.~\cite{Strikman:2005yv}. It was found that the rates are very large allowing one to probe deeply into the small-$x$ region. At NLO pQCD, the cross section of inclusive dijet photoproduction in Pb-Pb UPCs in the ATLAS kinematics
was calculated in~\cite{Guzey:2018dlm}. It was shown that the used theoretical framework provides a good description of various kinematic distributions measured by the ATLAS collaboration and that the calculated dijet photoproduction cross section is sensitive to nuclear modifications of nPDFs at the level of 10 to 20\%.

In this work, we explore the potential of inclusive dijet photoproduction in Pb-Pb UPCs in the LHC kinematics to provide new 
constraints on nPDFs. In particular, using the results of our NLO QCD calculations of the cross section of this process~\cite{Guzey:2018dlm} as pseudo-data, we 
study the effect of including these data using the Bayesian reweighting technique~\cite{Armesto:2013kqa,Paukkunen:2014zia,Kusina:2016fxy} on nCTEQ15, nCTEQ15np, and EPPS16 nPDFs. We find that depending on the assumed error on the pseudo-data, it leads to a significant reduction of uncertainties of nPDFs
at small $x_A$. For instance, taking the error to be 5\%, we find that the uncertainty of quark and gluon nCTEQ15 and nCTEQ15np nPDFs
reduces by approximately a factor of two at $x_A=10^{-3}$. The reweighting effect on EPPS16 nPDFs is much smaller due to the higher value of the tolerance and a more flexible parametrization form used in that analysis.

The remainder of this paper is structured as follows. In Sec.~\ref{sec:rew}, we summarize key steps of 
the Bayesian reweighting method and define our reweighting procedure. 
We present and discuss our results in Sec.~\ref{sec:results} and draw conclusions in Sec.~\ref{sec:conclusions}.

\section{Reweighting of the dijet photoproduction cross section}
\label{sec:rew}

To quantity the power of inclusive dijet photoproduction in Pb-Pb UPCs at the LHC to constrain nPDFs, we use the standard Bayesian reweighting procedure outlined in the literature~\cite{Armesto:2013kqa,Paukkunen:2014zia,Kusina:2016fxy}. In detail, 
starting with $2N$ error sets of nPDFs ($N=16$ for nCTEQ15~\cite{Kovarik:2015cma} and $N=20$ for EPPS16~\cite{Eskola:2016oht}), 
one generates $N_{\rm rep}=10,000$ replicas labeled by the index $k$ as follows
\begin{equation}
f_{j/A}^k(x,Q^2)=f_{j/A}^0(x,Q^2)+\frac{1}{2}\sum_{i=1}^{N} \left[f_{j/A}^{i+}(x,Q^2)-f_{j/A}^{i-}(x,Q^2)\right] R_{ki} \,,
\label{eq:replicas}
\end{equation}
where $f_{j/A}^0$ and $f_{j/A}^{i\pm}(x,Q^2)$ are the central fit and the plus and minus error nPDFs
corresponding to the eigenvector direction $i$ and $R_{ki}$ is a random number from the normal 
distribution centered at zero with the standard deviation of unity.

Next, for each PDF replica, one calculates the observable of interest, which in our case is 
the dijet photoproduction cross section as a function of $x_A$ ($x_A$ is the hadron-level estimate for the momentum fraction carried
by the interacting nuclear parton)~\cite{Guzey:2018dlm}:
\begin{equation}
\frac{d\sigma^k}{dx_A}  =  \sum_{a,b} \int^{y_{\rm max}}_{y_{\rm min}} dy \int^1_0 dx_{\gamma} f_{\gamma/A}(y)f_{a/\gamma}(x_{\gamma},\mu^2)f_{b/B}^k(x_A,\mu^2) d\hat{\sigma}(ab \to {\rm jets}) \,.
\label{eq:cs}
\end{equation}
In Eq.~(\ref{eq:cs}), $a,b$ are parton flavors; $f_{\gamma/A}(y)$ is the flux of equivalent photons
emitted by ion $A$, which depends on the photon light-cone momentum fraction $y$;
$f_{a/\gamma}(x_{\gamma},\mu^2)$ is the PDF of the photon for the resolved-photon contribution, which depends on the
momentum fraction $x_{\gamma}$ and the factorization scale $\mu$; $f_{b/B}(x_A,\mu^2)$
is the nuclear PDF with $x_A$ being the corresponding parton momentum fraction;
and $d\hat{\sigma}(ab \to {\rm jets})$ is the elementary cross section for production
of two- and three-parton final states emerging as jets in hard scattering of partons
$a$ and $b$. The sum over $a$ involves quarks and gluons for the resolved photon contribution 
and the photon for the direct photon contribution dominating at $x_{\gamma} \approx 1$.
The integration limits are determined by the rapidities and transverse momenta
 of the produced jets, see~\cite{Guzey:2018dlm} for details.
 Note that since the inclusive dijet cross section is linear in the nPDFs, which in turn are linear in $R_{ki}$, 
  it is sufficient to evaluate it $2N$ times for 
 each error nPDF.

The essence of the reweighting procedure is the calculation of statistical weights for each replica $w_k$, 
which quantify how well the calculation using Eq.~(\ref{eq:cs}) reproduces data or pseudo-data. In our case,
for the pseudo-data, we use the results of our calculation of the dijet cross section~\cite{Guzey:2018dlm}
\begin{equation}
\frac{d\sigma^0}{dx_A}  =  \sum_{a,b} \int^{y_{\rm max}}_{y_{\rm min}} dy \int^1_0 dx_{\gamma} f_{\gamma/A}(y)f_{a/\gamma}(x_{\gamma},\mu^2)f_{b/B}^0(x_A,\mu^2) d\hat{\sigma}(ab \to {\rm jets}) \,,
\label{eq:cs_0}
\end{equation}
where $f_{b/B}^0(x_A,\mu^2)$ corresponds to the central value of the nCTEQ15 nPDFs.
Then, the corresponding chi-squared $\chi_k^2$ is
\begin{equation}
\chi_k^2=\sum_{j=1}^{N_{\rm data}} \frac{(d\sigma^0/dx_A-d\sigma^k/dx_A)^2}{\sigma_j^2} \,,
\label{eq:chi}
\end{equation}
where the sum runs over the pseudo-data points;
$\sigma_j$ is the assumed uncertainty of the pseudo-data. In our case, $N_{\rm data}=9$ corresponding to different bins in $x_A$ and
we take $\sigma_j=\epsilon d\sigma^0/dx_A$ with $\epsilon=0.05$, 0.1,
0.15, and 0.2.
We assume that these errors account only for the statistical uncertainty and that bin-by-bin correlations are neglected.
The values of $\epsilon$ span the range of typical uncertainties of measurements of high-$E_T$ dijet photoproduction at HERA~\cite{Chekanov:2007ac}.

Finally, with the help of $\chi_k^2$, one can introduce the weights $w_k$ using the following relation:
\begin{equation}
w_k=\frac{e^{-\frac{1}{2}\chi_k^2/T}}{\frac{1}{N_{\rm rep}} \sum_i^{N_{\rm rep}}e^{-\frac{1}{2}\chi_i^2/T}} \,,
\label{eq:w}
\end{equation}
where $T$ is the tolerance associated with a given set of nPDFs.
 Note that $\sum_k w_k=N_{\rm rep}$. In our analysis, we use
$T=35$ for nCTEQ15 and nCTEQ15np~\cite{Kovarik:2015cma} and $T=52$ for EPPS16~\cite{Eskola:2016oht}.

Using the weights $w_k$, one can calculate the new, weighted average cross section and its error: 
\begin{eqnarray}
\left\langle \frac{d\sigma}{dx_A} \right\rangle_{\rm new} &=& \frac{1}{N_{\rm rep}} \sum_{k=1}^{N_{\rm rep}} w_k \frac{d\sigma^k}{dx_A}\,, \nonumber \\
\delta \left\langle \frac{d\sigma}{dx_A} \right\rangle_{\rm new} &=& \sqrt{\frac{1}{N_{\rm rep}} 
\sum_k w_k\left(\frac{d\sigma^k}{dx_A}-\left\langle \frac{d\sigma}{dx_A} \right\rangle_{\rm new}\right)^2} \,.
\label{eq:cs_new}
\end{eqnarray}

Similarly, one can evaluate the reweighted nPDFs and their uncertainties:
\begin{eqnarray}
\langle f_{j/A}(x,Q^2) \rangle_{\rm new} &=& \frac{1}{N_{\rm rep}} \sum_{k=1}^{N_{\rm rep}} w_k f_{j/A}^k(x,Q^2) \,, \nonumber \\
\delta \langle f_{j/A}(x,Q^2) \rangle_{\rm new} &=& \sqrt{\frac{1}{N_{\rm rep}} \sum_{k=1}^{N_{\rm rep}} 
w_k \left(f_{j/A}^k-\langle f_{j/A}(x,Q^2) \rangle_{\rm new}\right)^2} \,.
\label{eq:average_new}
\end{eqnarray}
Equations~(\ref{eq:cs_new}) and (\ref{eq:average_new}) quantify the effect of the pseudo-data on the calculation of
the cross section of inclusive dijet photoproduction 
and the central value and uncertainties of nPDFs, respectively.

The effective number of replicas contributing to Eqs.~(\ref{eq:cs_new}) and (\ref{eq:average_new})
can be estimated using the following expression:
\begin{equation}
N_{\rm eff}=\exp \left[\frac{1}{N_{\rm rep}} \sum_k^{N_{\rm rep}} w_k \ln(N_{\rm rep}/w_k) \right] \,.
\label{eq:N_eff} 
\end{equation}
Table~\ref{table:N_eff} summarizes our values of $N_{\rm eff}$ for $\epsilon=0.05$, 0.1, 0.15, and 0.2 and nCTEQ15, nCTEQ15np, and EPPS16 nPDFs.
\begin{table}[h]
\caption{The effective number of replicas $N_{\rm eff}$ for different choices of the experimental error and sets of nPDFs.}
\begin{center}
\begin{tabular}{|c|c|c|c|}
\hline
$\epsilon$ & $N_{\rm eff}$(nCTEQ15) &  $N_{\rm eff}$(nCTEQ15np) &  $N_{\rm eff}$(EPPS16) \\
\hline
0.05 & 4407 & 3982  & 5982 \\
0.1 &  7483 & 7742  & 8727 \\
0.15 & 8870 & 9107  & 9555 \\
0.2  & 9464 & 9607  & 9818 \\
\hline
\end{tabular}
\end{center}
\label{table:N_eff}
\end{table}%

\section{Results of the reweighting}
\label{sec:results}

\begin{figure}[t]
\begin{center}
\epsfig{file=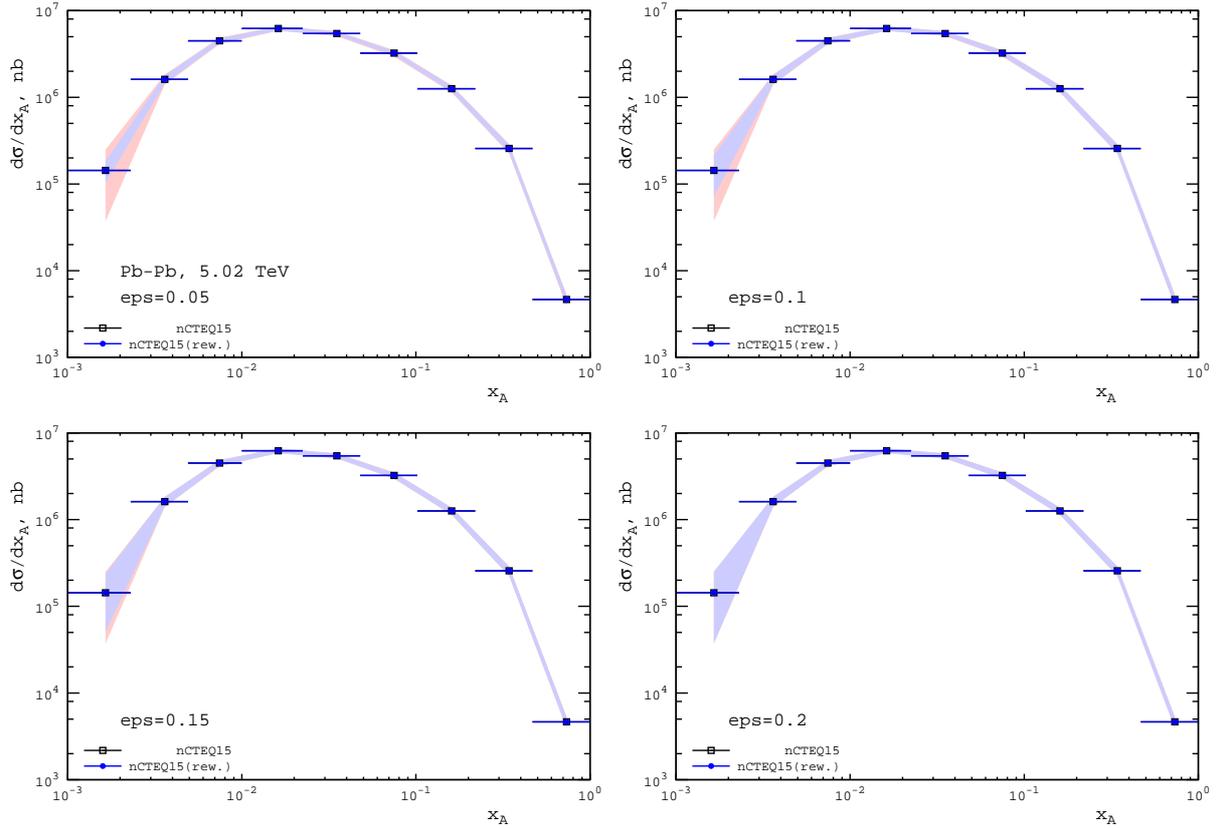,scale=0.7}
 \caption{The dijet photoproduction cross section as a function of $x_A$ with (blue solid circles and error bands) and without 
 (black open squares and red error bands) the Bayesian reweighting.
 The calculations correspond to the nCTEQ15 nPDFs. Different panels show the results for the four considered cases of the assumed
 error $\epsilon=0.05$, 0.1, 0.15, and 0.2.}
 \label{fig:data_reweighting_new}
\end{center}
\end{figure}

\begin{figure}[t]
\begin{center}
\epsfig{file=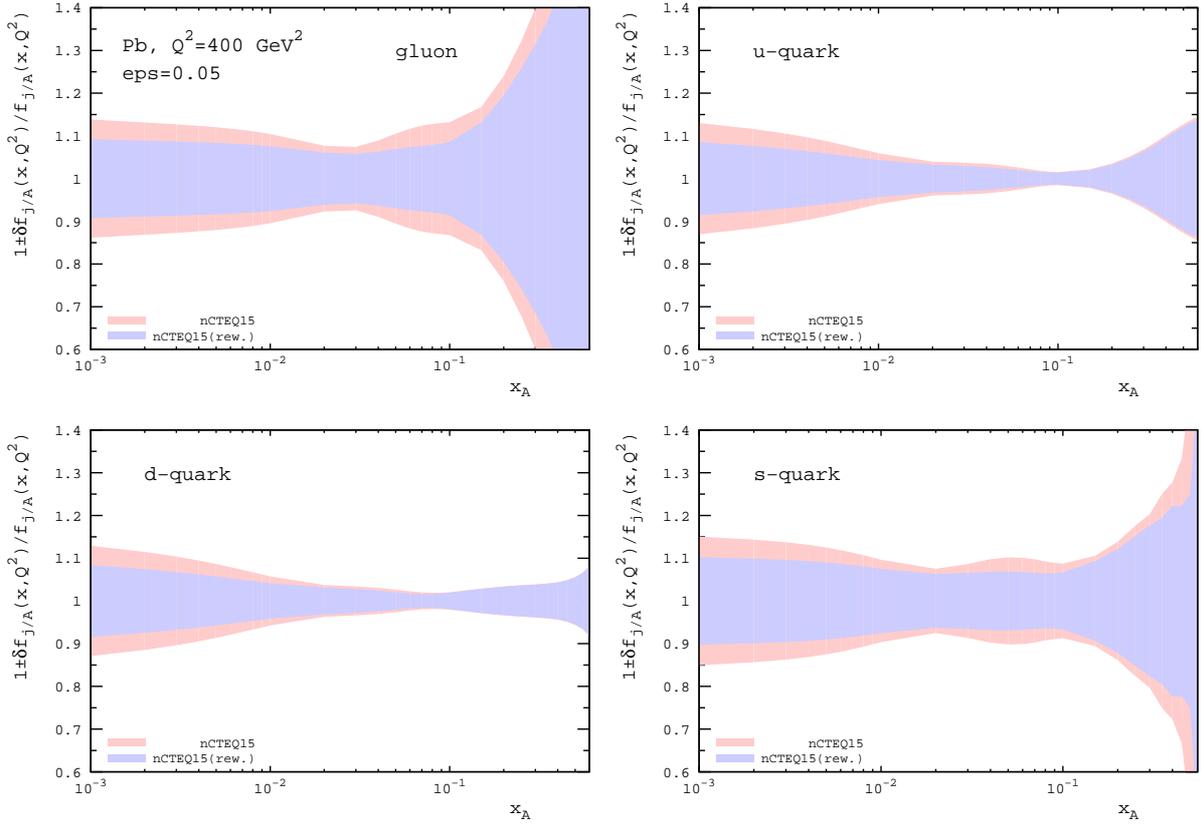,scale=0.7}
 \caption{The gluon, $u$-quark, $d$-quark, and $s$-quark nCTEQ15 nPDFs as a function of $x_A$ at $Q^2=400$ GeV$^2$ 
 with (blue, inner band) and without (red, outer band) the Bayesian reweighting. The case of $\epsilon=0.05$.
 }
 \label{fig:pdfs_reweighting_new_005}
\end{center}
\end{figure}

\begin{figure}[t]
\begin{center}
\epsfig{file=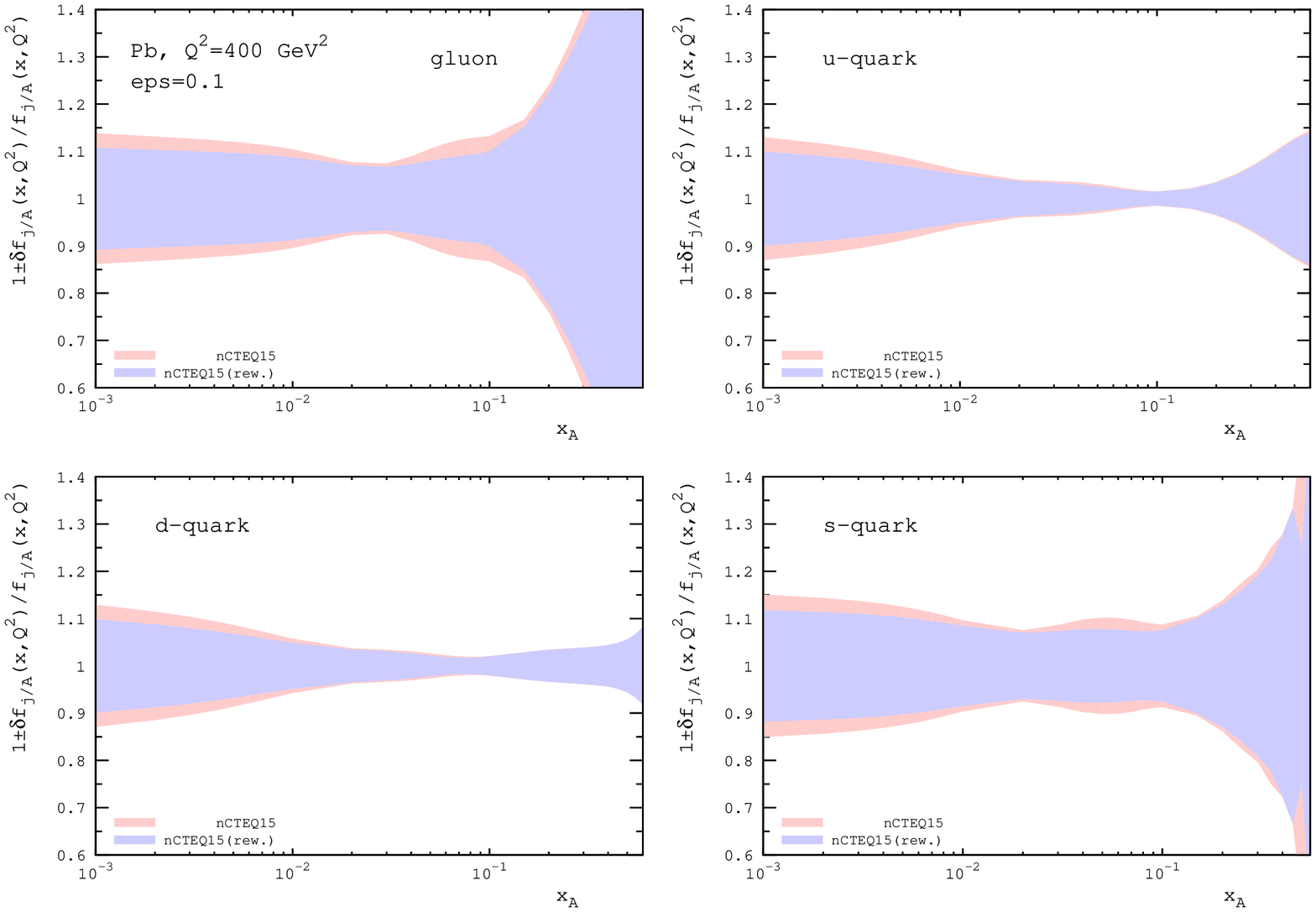,scale=0.7}
 \caption{The same as Fig.~\ref{fig:pdfs_reweighting_new_01}, but with $\epsilon=0.1$.
 }
\label{fig:pdfs_reweighting_new_01}
\end{center}
\end{figure}

\begin{figure}[t]
\begin{center}
\epsfig{file=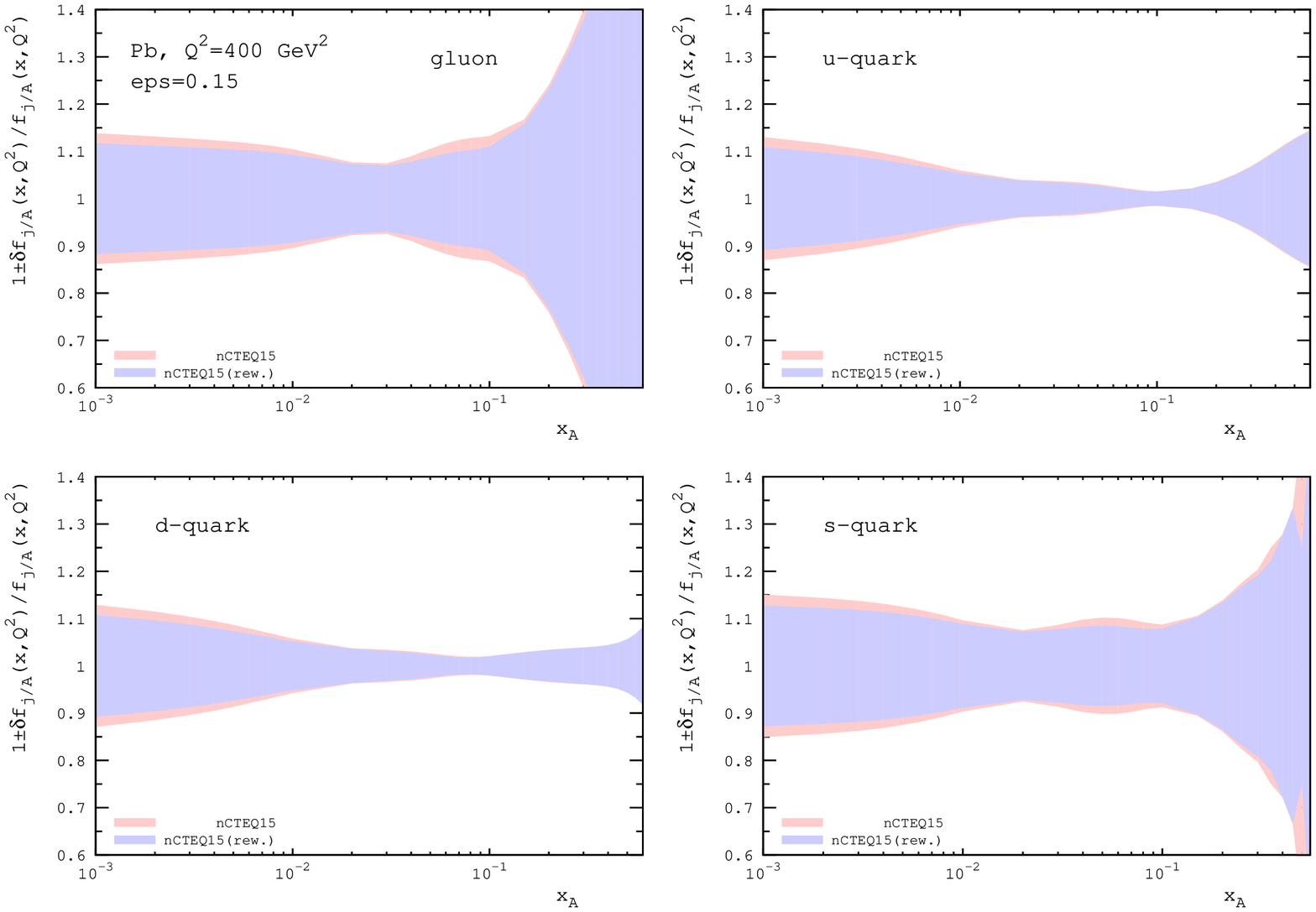,scale=0.7}
 \caption{The same as Fig.~\ref{fig:pdfs_reweighting_new_01}, but with $\epsilon=0.15$.
 }
\label{fig:pdfs_reweighting_new_015}
\end{center}
\end{figure}

\begin{figure}[t]
\begin{center}
\epsfig{file=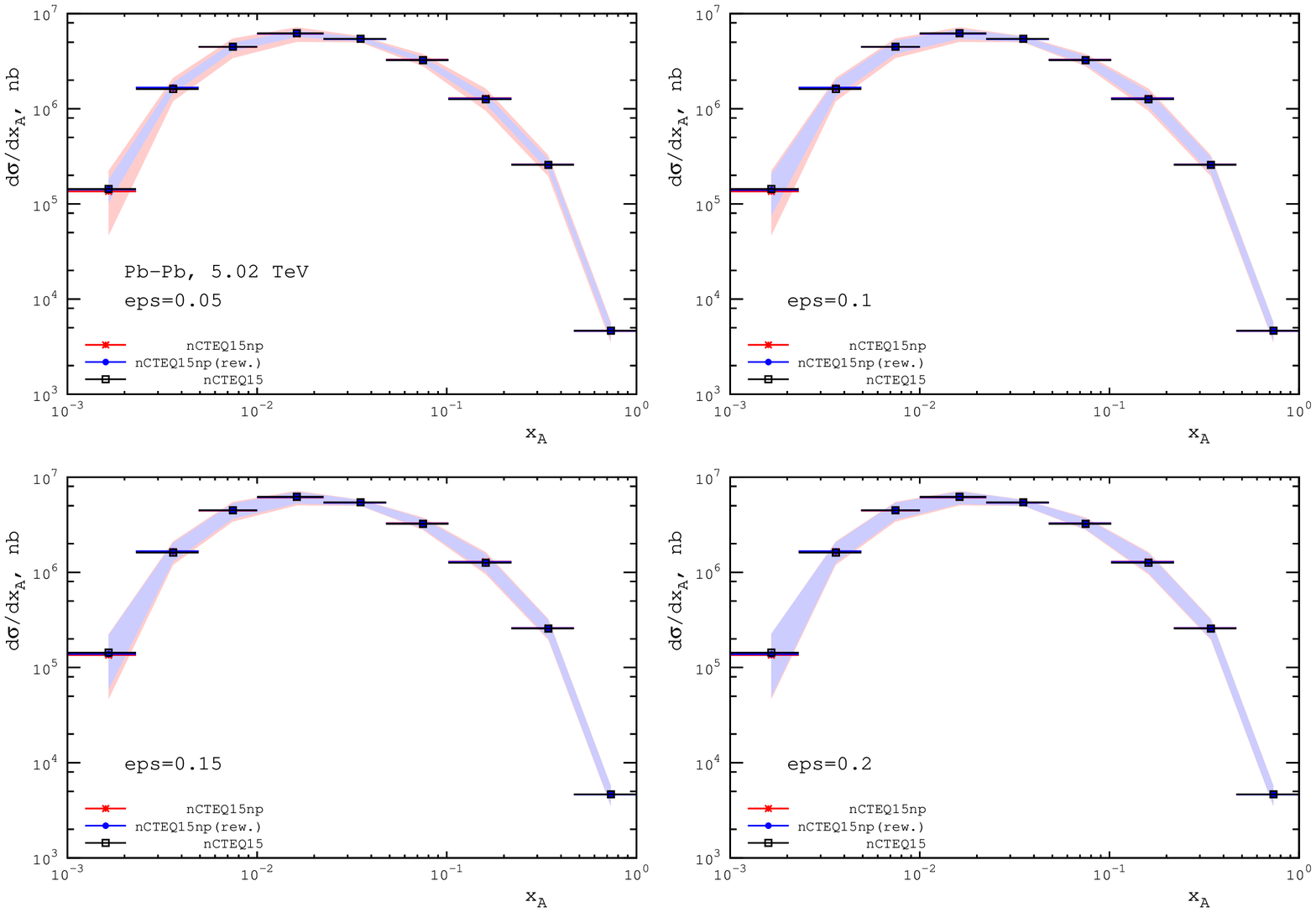,scale=0.7}
 \caption{The dijet photoproduction cross section as a function of $x_A$ with (blue solid circles and error bands) 
 and without (red crosses and error bands) the Bayesian reweighting calculated using 
 the nCTEQ15np nPDFs; the cross section used as pseudo-data is calculated with nCTEQ15 (open black squares).
  Different panels show the results for the four considered cases of the assumed
 error $\epsilon$.
 }
 \label{fig:data_reweighting_np_new}
\end{center}
\end{figure}

\begin{figure}[t]
\begin{center}
\epsfig{file=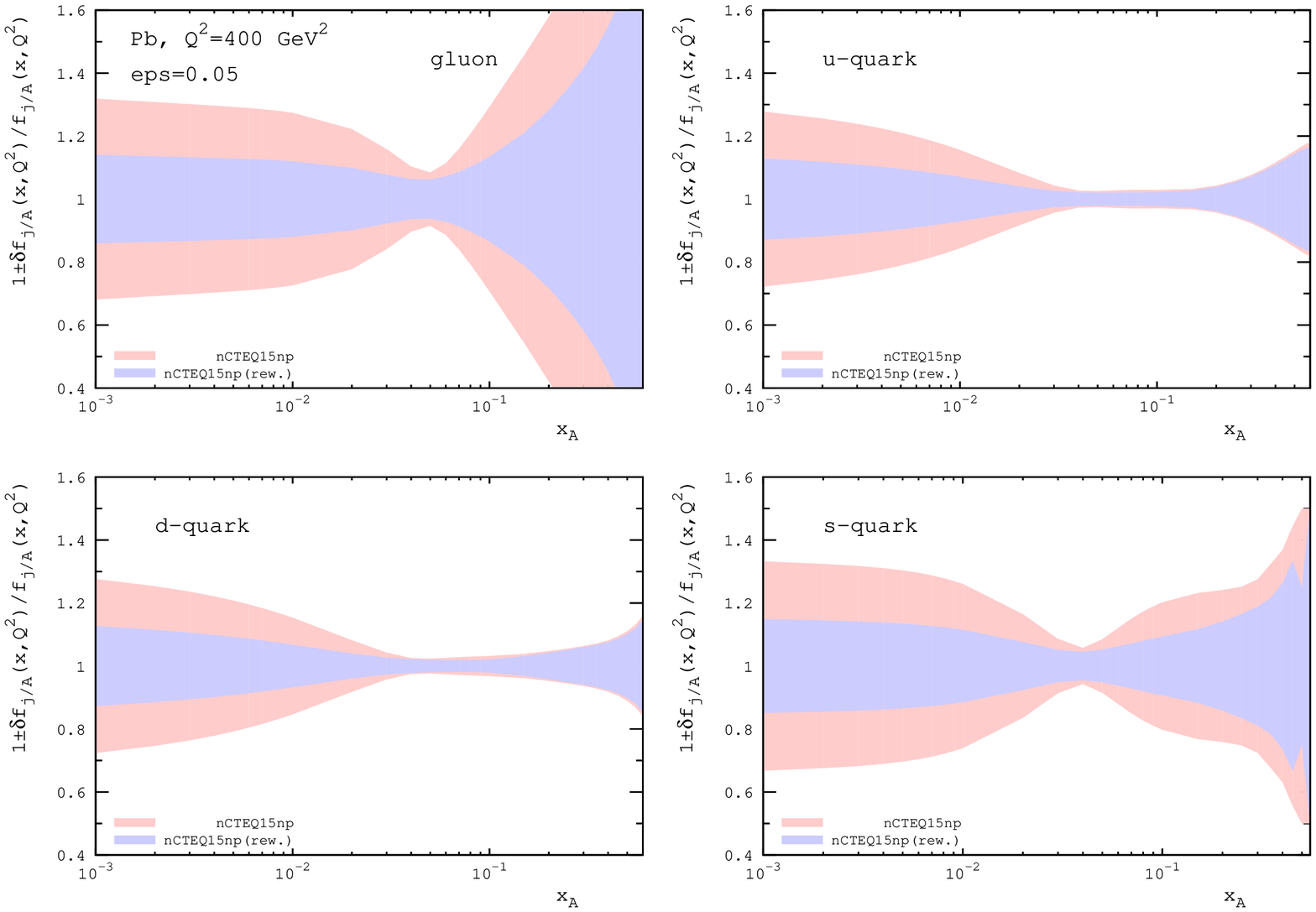,scale=0.7}
 \caption{The gluon, $u$-quark, $d$-quark, and $s$-quark nCTEQnp nPDFs as a function of $x$ at $Q^2=400$ GeV$^2$ 
 with (blue, inner band) and without (red, outer band) the Bayesian reweighting. The case of $\epsilon=0.05$.}
\label{fig:pdfs_reweighting_np_new_005}
\end{center}
\end{figure}

\begin{figure}[t]
\begin{center}
\epsfig{file=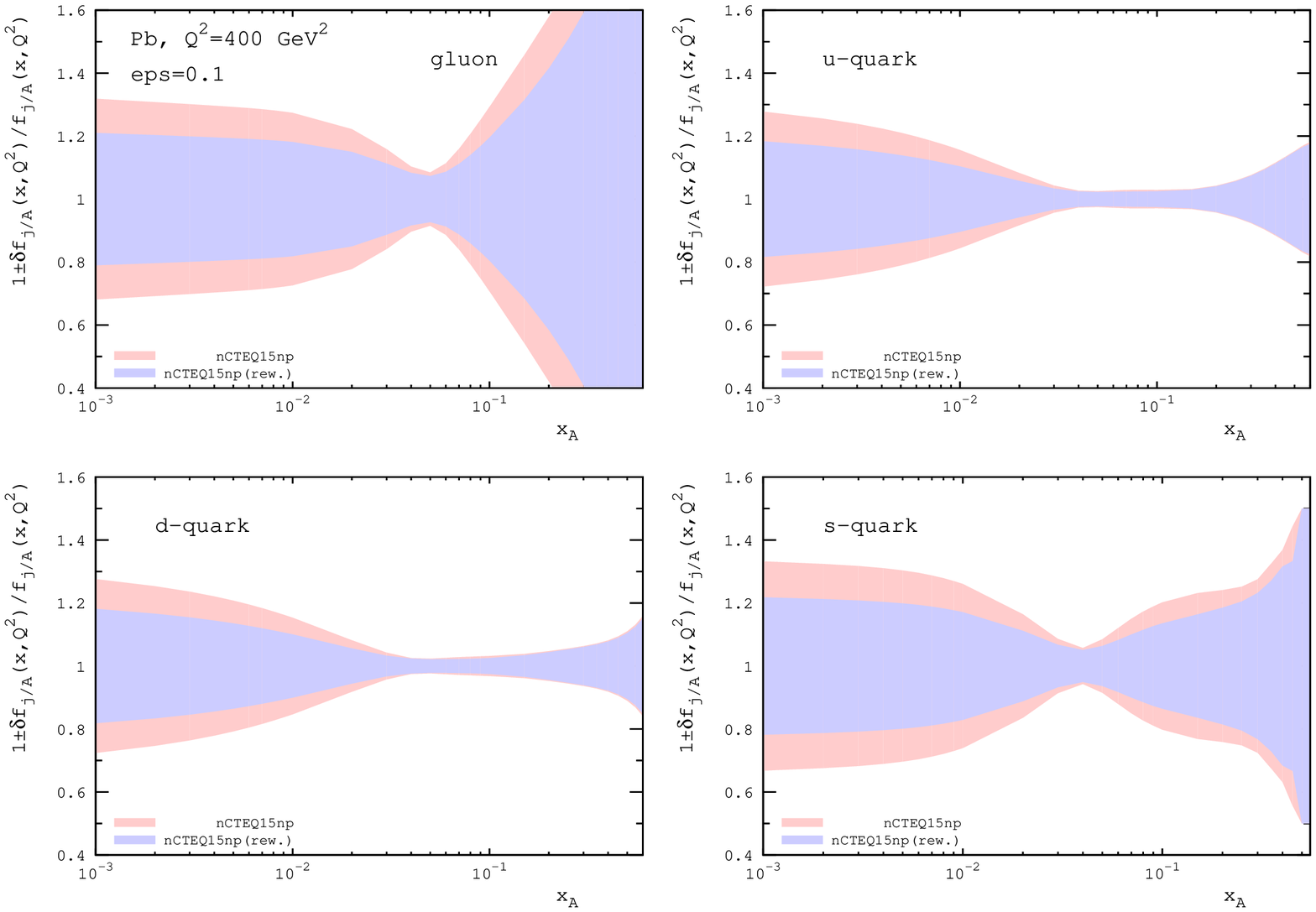,scale=0.7}
 \caption{The same as Fig.~\ref{fig:pdfs_reweighting_np_new_005}, but with $\epsilon=0.1$.}
\label{fig:pdfs_reweighting_np_new_01}
\end{center}
\end{figure}

\begin{figure}[t]
\begin{center}
\epsfig{file=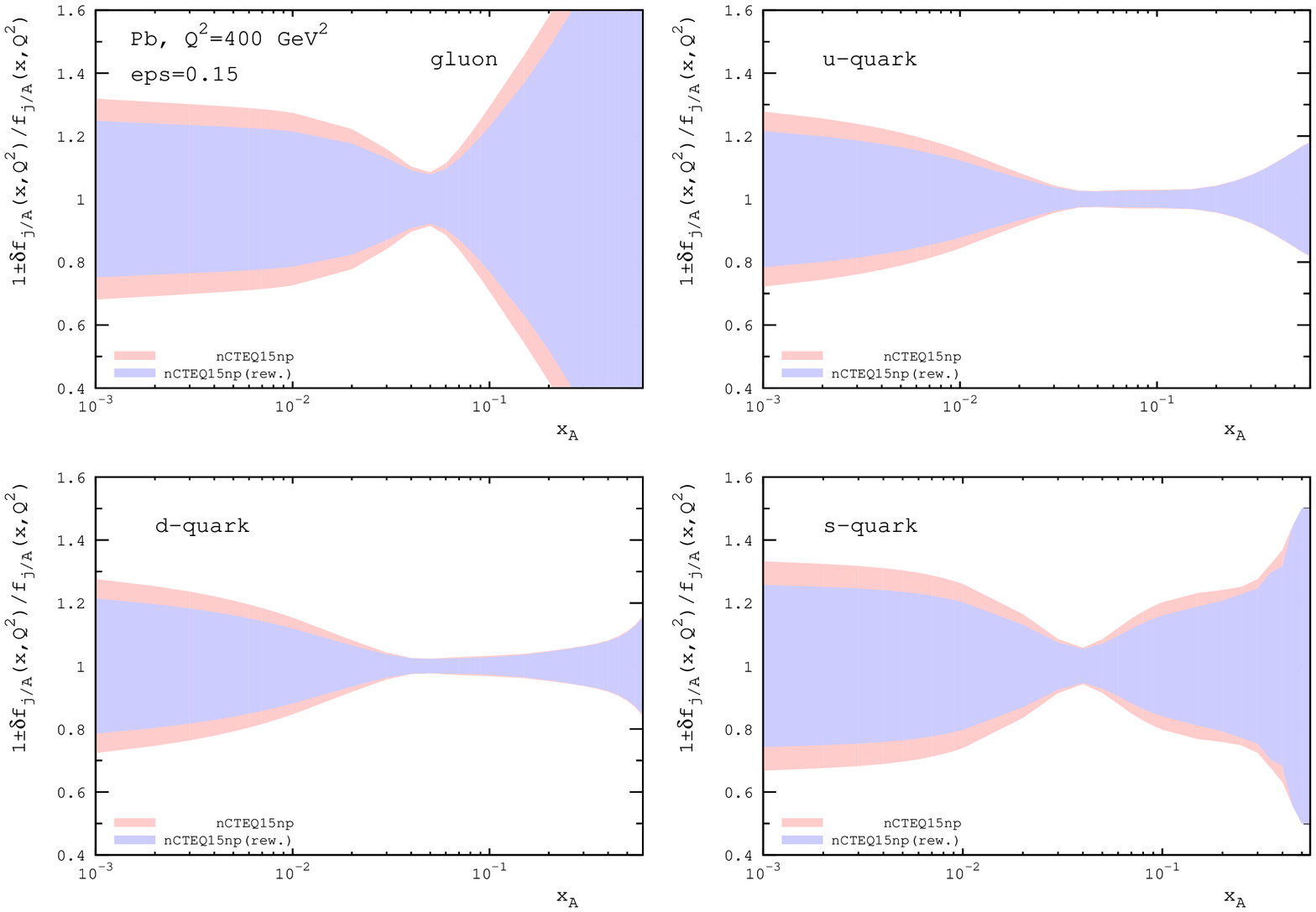,scale=0.7}
 \caption{The same as Fig.~\ref{fig:pdfs_reweighting_np_new_005}, but with $\epsilon=0.15$.}
\label{fig:pdfs_reweighting_np_new_015}
\end{center}
\end{figure}

\begin{figure}[t]
\begin{center}
\epsfig{file=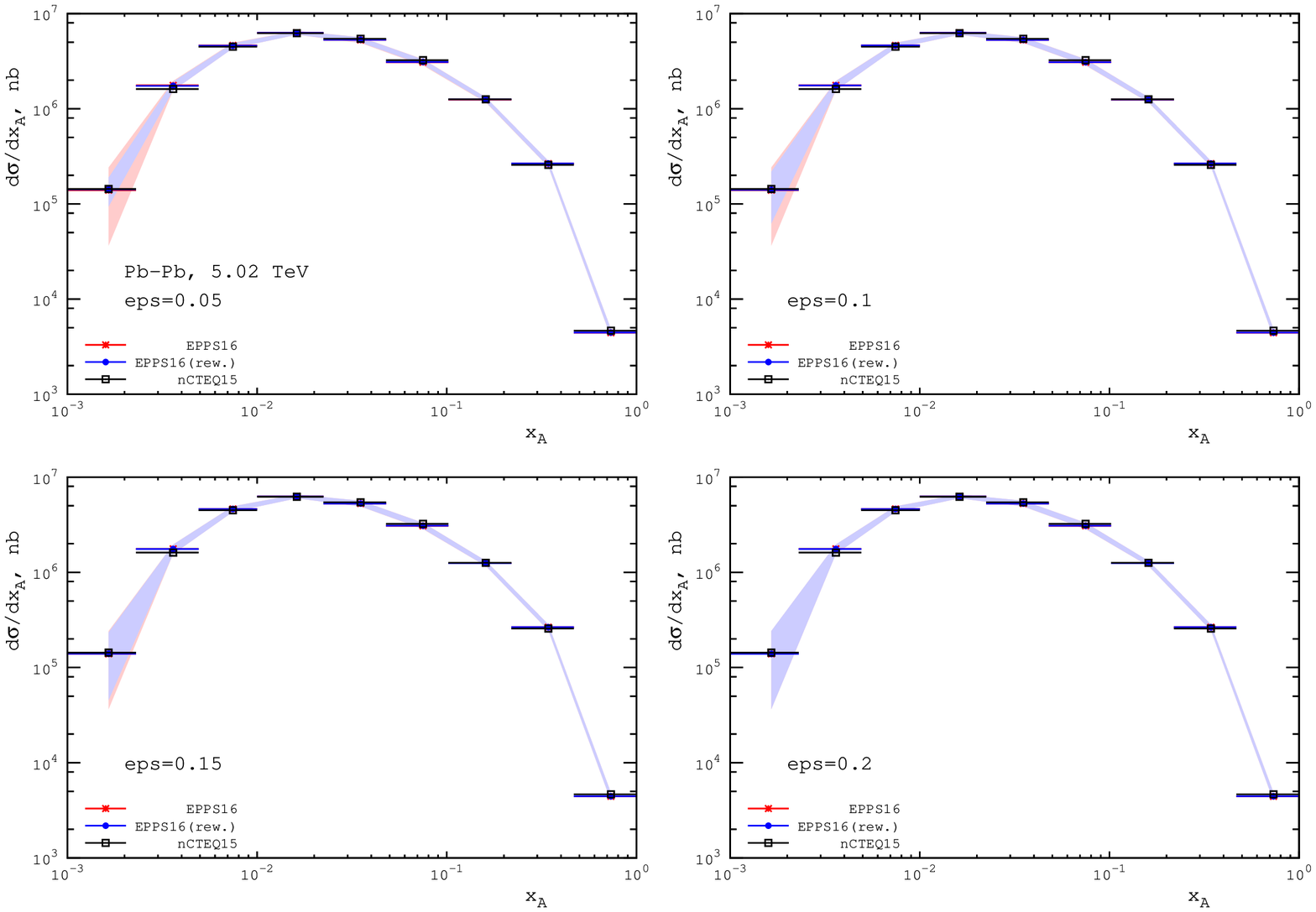,scale=0.7}
 \caption{The dijet photoproduction cross section as a function of $x_A$ with (blue solid circles and error bands) and without (red crosses and error bands) the Bayesian reweighting calculated using the EPPS16 nPDFs; the cross section used as pseudo-data is calculated with nCTEQ15 (open black squares).
  Different panels show the results for the four considered cases of the assumed
 error $\epsilon$.}
 \label{fig:data_reweighting_epps16_new}
\end{center}
\end{figure}

\begin{figure}[t]
\begin{center}
\epsfig{file=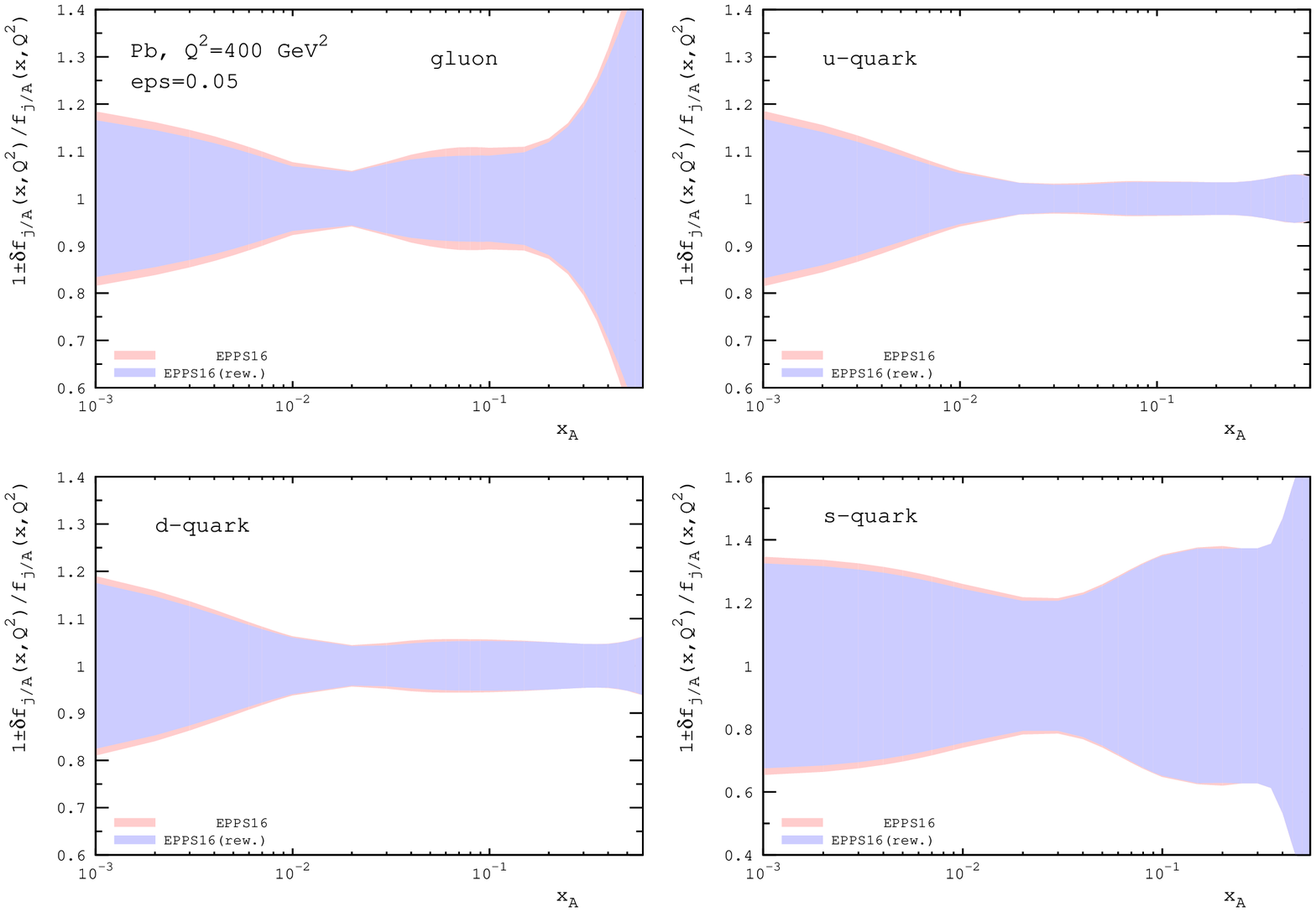,scale=0.7}
 \caption{The gluon, $u$-quark, $d$-quark, and $s$-quark EPPS16 nPDFs as a function of $x$ at $Q^2=400$ GeV$^2$ 
 with (blue, inner band) and without (red, outer band) the Bayesian reweighting. The case of $\epsilon=0.05$.}
\label{fig:pdfs_reweighting_epps16_new_005}
\end{center}
\end{figure}

\begin{figure}[t]
\begin{center}
\epsfig{file=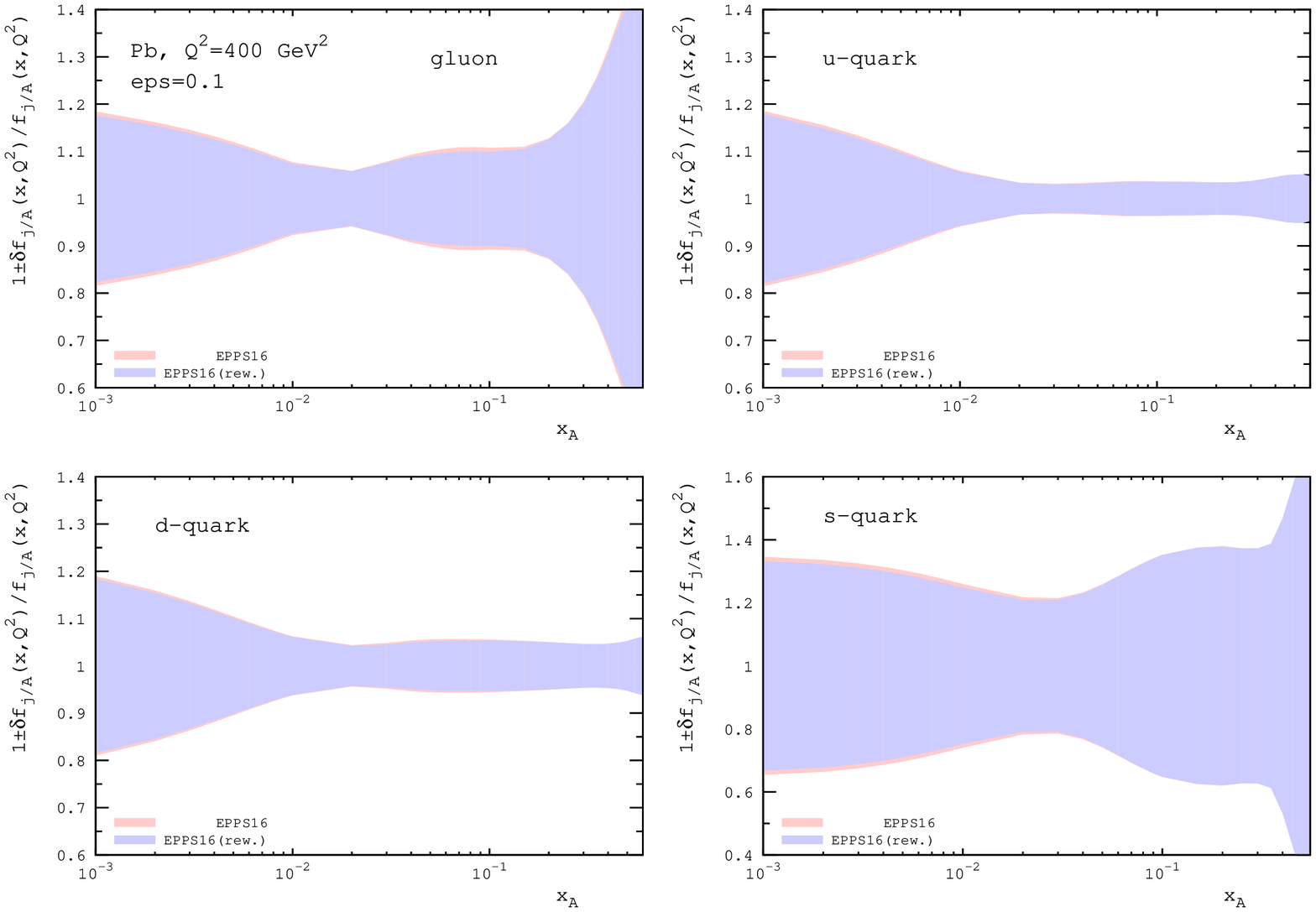,scale=0.7}
 \caption{The same as Fig.~\ref{fig:pdfs_reweighting_epps16_new_005}, but with $\epsilon=0.1$.}
\label{fig:pdfs_reweighting_epps16_new_01}
\end{center}
\end{figure}

Using the procedure outlined in Sec.~\ref{sec:rew}, we perform the Bayesian reweighting of the
pseudo-data on the cross section of inclusive dijet photoproduction on nuclei in Pb-Pb UPCs in the LHC kinematics. Our results are shown in Figs.~\ref{fig:data_reweighting_new}--\ref{fig:pdfs_reweighting_new_015} for nCTEQ15 nPDFs,
Figs.~\ref{fig:data_reweighting_np_new}--\ref{fig:pdfs_reweighting_np_new_015} for nCTEQ15np nPDFs, and 
Figs.~\ref{fig:data_reweighting_epps16_new}--\ref{fig:pdfs_reweighting_epps16_new_01} for EPPS16 nPDFs.

Figures~\ref{fig:data_reweighting_new}, \ref{fig:data_reweighting_np_new}, and \ref{fig:data_reweighting_epps16_new} 
show the dijet cross section as a function of $x_A$: the pseudo-data points labeled ``nCTEQ15'' and given by black open squares
are the results of the calculation using the central nCTEQ15 fit; 
red crosses with the associated error bands are
the results of the calculations using a given set of nPDFs (the crosses coincide with the open squares in Fig.~\ref{fig:data_reweighting_new}, and, hence, are not
shown); 
finally, the blue filled circles and the associated error bands show the reweighted cross section and its uncertainty, see 
Eq.~(\ref{eq:cs_new}). The four panels correspond to our four choices of the assumed error $\epsilon$.
One can see from these figures that while the reweighting does not noticeably change the central values of the cross section, 
it reduces its theoretical uncertainty: the effect is largest for the smallest $\epsilon$ and the first small-$x_A$ bin.
\footnote{This bin has larger statistical uncertainties, which can however be reduced by increasing the precision of the Monte Carlo integration \cite{Guzey:2018dlm}.}

The remaining figures (Figs.~\ref{fig:pdfs_reweighting_new_005}--\ref{fig:pdfs_reweighting_new_015}, 
\ref{fig:pdfs_reweighting_np_new_005}--\ref{fig:pdfs_reweighting_np_new_015}, and \ref{fig:pdfs_reweighting_epps16_new_005}--\ref{fig:pdfs_reweighting_epps16_new_01}) demonstrate the effect of the reweighting on 
uncertainties of nPDFs: 
different panels show 
uncertainty bands of nPDFs normalized to their central value, i.e., the bands spanned by $1 \pm \delta f_{j/A}(x,Q^2)/f_{j/A}(x,Q^2)$, see Eq.~(\ref{eq:average_new}),
for the gluon, $u$-quark, $d$-quark, and $s$-quark nPDFs before (red, outer band) and after (blue, inner band) the 
reweighting as a function of the momentum fraction $x_A$ at $Q^2=400$ GeV$^2$. This is a characteristic value of $Q^2$ probed
in dijet photoproduction in the ATLAS kinematics.
While the central values of nPDFs are essentially not affected by the reweighting, the uncertainty bands for nCTEQ15 and 
nCTEQ15np are noticeably reduced.
As expected, the effect is largest at $\epsilon=0.05$ and much smaller at 
$\epsilon=0.15$ and $\epsilon=0.2$.
(Since the reduction of the uncertainty bands is very similar in the $\epsilon=0.15$ and $\epsilon=0.2$ cases, we only 
show the results for the former.) 
For instance, the uncertainty in the gluon and quark nPDFs at $x_A=0.001$ reduces by approximately a factor of two.
It is interesting to note that the uncertainty of the small-$x_A$ gluon distribution in the case of nCTEQ15np after
the reweighting is similar to that of nCTEQ15 before the reweighting -- it is of the order of 15\% in both cases. Therefore, 
dijet photoproduction should have a similar impact on nCTEQ15 nPDFs as the RHIC inclusive pion production data,
which was included in nCTEQ15 and excluded in nCTEQ15np. 
The advantage of dijet photoproduction is that it does not involve the pion fragmentation functions, which 
necessarily brings an additional uncertainty in analyses of nPDFs.

In the case of EPPS16 nPDFs (see Figs.~\ref{fig:pdfs_reweighting_epps16_new_005} and \ref{fig:pdfs_reweighting_epps16_new_01}), 
the effect of reweighting is much smaller due to several reasons. First, these nPDFs have been obtained with a higher value of the
tolerance $T$, which allows for significantly more replicas to contribute to the reweighted quantities 
(see Table~\ref{table:N_eff}) and which reduces the effectiveness of the reweighting.
Second, a more flexible form of the EPPS16 nPDF parametrization also significantly reduces the reweighting effect, which is 
negligibly small in the $\epsilon=0.15$ and $\epsilon=0.2$ cases. We therefore do not show them here, since the blue and red solid lines and 
bands completely overlap.

Note that in typical fits of nPDFs, one parametrizes the dependence of the fit parameters on the
nuclear mass number $A$~\cite{Eskola:2009uj,Kovarik:2015cma,Eskola:2016oht}, which hence correlates these parameters for different nuclei. While, by construction, nuclear modifications of nPDFs and their uncertainties decrease with a decrease of $A$,
the reduction of uncertainties of nPDFs for Pb due to the considered reweighting should also reduce uncertainties of 
nPDFs for lighter nuclei; the magnitude of the effect depends on numerical values of the fit parameters.

\section{Conclusions}
\label{sec:conclusions}

In this work, we studied the potential of inclusive dijet photoproduction in Pb-Pb UPCs in the LHC kinematics to give new 
constraints on nPDFs. Using the results of our NLO QCD calculations of the cross section of this process as pseudo-data, we 
analyzed the effect of including these data using the Bayesian reweighting technique on nCTEQ15, nCTEQ15np, and EPPS16 nPDFs. 
We found that depending on the assumed error on the pseudo-data, it leads to a significant reduction of the nPDF uncertainties
at small $x_A$. For instance, taking the error to be 5\%, we find that the uncertainty of quark and gluon nCTEQ15 and nCTEQ15np nPDFs
reduces by a factor of two at $x_A=10^{-3}$. 
We observed that the uncertainty of the small-$x_A$ gluon distribution in the nCTEQ15np case after
the reweighting is similar to that of nCTEQ15 before the reweighting, which indicates that 
dijet photoproduction should have a similar impact on nCTEQ15 nPDFs as the RHIC inclusive pion production data with 
the advantage that dijet photoproduction is free from the uncertainty associated with the pion fragmentation functions.
At the same time, the reweighting effect on EPPS16 nPDFs is much smaller due to the higher value of the tolerance and a more flexible parametrization form used in the EPPS16 analysis.

\begin{acknowledgements}

VG would like to thank K.~Kovarik and H.~Paukkunen for useful discussions. 
VG's research is supported in part by 
RFBR, research project 17-52-12070. The authors gratefully acknowledge financial 
support of DFG through the grant KL 1266/9-1 within the framework of the joint German-Russian project ``New constraints on nuclear parton distribution 
functions at small $x$ from dijet production in $\gamma A$ collisions at the LHC".

\end{acknowledgements}


\begin{thebibliography}{99}

\bibitem{deFlorian:2003qf}
  D.~de Florian and R.~Sassot,
  Phys.\ Rev.\ D {\bf 69} (2004) 074028
  [hep-ph/0311227].

\bibitem{Hirai:2007sx}
  M.~Hirai, S.~Kumano and T.-H.~Nagai,
  Phys.\ Rev.\ C {\bf 76} (2007) 065207
  [arXiv:0709.3038 [hep-ph]].

\bibitem{Eskola:2009uj}
  K.~J.~Eskola, H.~Paukkunen and C.~A.~Salgado,
  JHEP {\bf 0904} (2009) 065
  [arXiv:0902.4154 [hep-ph]].

\bibitem{deFlorian:2011fp}
  D.~de Florian, R.~Sassot, P.~Zurita and M.~Stratmann,
  Phys.\ Rev.\ D {\bf 85} (2012) 074028
  [arXiv:1112.6324 [hep-ph]].

\bibitem{Kovarik:2015cma}
  K.~Kovarik {\it et al.},
  Phys.\ Rev.\ D {\bf 93} (2016) no.8,  085037
  [arXiv:1509.00792 [hep-ph]].

\bibitem{Khanpour:2016pph}
  H.~Khanpour and S.~Atashbar Tehrani,
  Phys.\ Rev.\ D {\bf 93} (2016) no.1,  014026
  [arXiv:1601.00939 [hep-ph]].


\bibitem{Eskola:2016oht}
  K.~J.~Eskola, P.~Paakkinen, H.~Paukkunen and C.~A.~Salgado,
  Eur.\ Phys.\ J.\ C {\bf 77} (2017) no.3,  163
  [arXiv:1612.05741 [hep-ph]].

\bibitem{Salgado:2011wc}
  C.~A.~Salgado {\it et al.},
  J.\ Phys.\ G {\bf 39} (2012) 015010
  [arXiv:1105.3919 [hep-ph]].

\bibitem{Paukkunen:2018kmm}
  H.~Paukkunen,
  arXiv:1811.01976 [hep-ph].

\bibitem{Citron:2018lsq}
  Z.~Citron {\it et al.},
  arXiv:1812.06772 [hep-ph].

\bibitem{Accardi:2012qut}
  A.~Accardi {\it et al.},
  Eur.\ Phys.\ J.\ A {\bf 52} (2016) no.9,  268
  [arXiv:1212.1701 [nucl-ex]].

\bibitem{Aschenauer:2017oxs}
  E.~C.~Aschenauer, S.~Fazio, M.~A.~C.~Lamont, H.~Paukkunen and P.~Zurita,
  Phys.\ Rev.\ D {\bf 96} (2017) no.11,  114005
  [arXiv:1708.05654 [nucl-ex]].


\bibitem{AbelleiraFernandez:2012cc}
  J.~L.~Abelleira Fernandez {\it et al.} [LHeC Study Group],
  J.\ Phys.\ G {\bf 39} (2012) 075001
  [arXiv:1206.2913 [physics.acc-ph]].


\bibitem{Eskola:2013aya}
  K.~J.~Eskola, H.~Paukkunen and C.~A.~Salgado,
  JHEP {\bf 1310} (2013) 213
  [arXiv:1308.6733 [hep-ph]].

\bibitem{Helenius:2014qla}
  I.~Helenius, K.~J.~Eskola and H.~Paukkunen,
  JHEP {\bf 1409} (2014) 138
  [arXiv:1406.1689 [hep-ph]].

\bibitem{Armesto:2015lrg}
  N.~Armesto, H.~Paukkunen, J.~M.~Penin, C.~A.~Salgado and P.~Zurita,
  Eur.\ Phys.\ J.\ C {\bf 76} (2016) no.4,  218
  [arXiv:1512.01528 [hep-ph]].

\bibitem{dEnterria:2015mgr}
  D.~d'Enterria, K.~Krajcz\'ar and H.~Paukkunen,
  Phys.\ Lett.\ B {\bf 746} (2015) 64
  [arXiv:1501.05879 [hep-ph]].

\bibitem{Brandt:2014vva}
  M.~Brandt, M.~Klasen and F.~K\"onig,
  Nucl.\ Phys.\ A {\bf 927} (2014) 78
  [arXiv:1401.6817 [hep-ph]].

\bibitem{Eskola:2018sxu}
  K.~J.~Eskola, P.~Paakkinen and H.~Paukkunen,
  arXiv:1812.05438 [hep-ph].

\bibitem{Baltz:2007kq}
  A.~J.~Baltz {\it et al.},
  Phys.\ Rept.\  {\bf 458} (2008) 1
  [arXiv:0706.3356 [nucl-ex]].

\bibitem{Abbas:2013oua}
  E.~Abbas {\it et al.} [ALICE Collaboration],
  Eur.\ Phys.\ J.\ C {\bf 73} (2013) no.11,  2617
  [arXiv:1305.1467 [nucl-ex]].

\bibitem{Abelev:2012ba}
  B.~Abelev {\it et al.} [ALICE Collaboration],
  Phys.\ Lett.\ B {\bf 718} (2013) 1273
  [arXiv:1209.3715 [nucl-ex]].

\bibitem{Adam:2015sia}
  J.~Adam {\it et al.} [ALICE Collaboration],
  Phys.\ Lett.\ B {\bf 751} (2015) 358
  [arXiv:1508.05076 [nucl-ex]].

\bibitem{Khachatryan:2016qhq} 
  V.~Khachatryan {\it et al.} [CMS Collaboration],
  Phys.\ Lett.\ B {\bf 772} (2017) 489
  [arXiv:1605.06966 [nucl-ex]].

\bibitem{Ryskin:1992ui}
  M.~G.~Ryskin,
  Z.\ Phys.\ C {\bf 57} (1993) 89.

\bibitem{Guzey:2013xba} 
  V.~Guzey, E.~Kryshen, M.~Strikman and M.~Zhalov,
  Phys.\ Lett.\ B {\bf 726} (2013) 290
  [arXiv:1305.1724 [hep-ph]].
  
\bibitem{Guzey:2013qza} 
  V.~Guzey and M.~Zhalov,
  JHEP {\bf 1310} (2013) 207
  [arXiv:1307.4526 [hep-ph]].
  
\bibitem{Frankfurt:2011cs}
  L.~Frankfurt, V.~Guzey and M.~Strikman,
  Phys.\ Rept.\  {\bf 512} (2012) 255
  [arXiv:1106.2091 [hep-ph]].
  
\bibitem{Ivanov:2004vd}
  D.~Y.~Ivanov, A.~Schafer, L.~Szymanowski and G.~Krasnikov,
  Eur.\ Phys.\ J.\ C {\bf 34} (2004) no.3,  297
   Erratum: [Eur.\ Phys.\ J.\ C {\bf 75} (2015) no.2,  75]
  [hep-ph/0401131].

\bibitem{Jones:2015nna}
  S.~P.~Jones, A.~D.~Martin, M.~G.~Ryskin and T.~Teubner,
  J.\ Phys.\ G {\bf 43} (2016) no.3,  035002
  [arXiv:1507.06942 [hep-ph]].

\bibitem{Klein:2017vua}
  S.~R.~Klein,
  Nucl.\ Phys.\ A {\bf 967} (2017) 249
  [arXiv:1704.04715 [nucl-ex]].
  
\bibitem{Kryshen:2017jfz}
  E.~L.~Kryshen [ALICE Collaboration],
  Nucl.\ Phys.\ A {\bf 967} (2017) 273
  [arXiv:1705.06872 [nucl-ex]].
  
\bibitem{Guzey:2016piu}
  V.~Guzey, E.~Kryshen and M.~Zhalov,
  Phys.\ Rev.\ C {\bf 93} (2016) no.5,  055206
  [arXiv:1602.01456 [nucl-th]].

\bibitem{Atlas} The ATLAS collaboration, "Photo-nuclear dijet production in ultra-peripheral Pb+Pb collisions",
ATLAS-CONF-2017-011, February 2017.

\bibitem{Strikman:2005yv}
  M.~Strikman, R.~Vogt and S.~N.~White,
  Phys.\ Rev.\ Lett.\  {\bf 96} (2006) 082001
  [hep-ph/0508296].

\bibitem{Guzey:2018dlm}
  V.~Guzey and M.~Klasen,
  arXiv:1811.10236 [hep-ph].

\bibitem{Armesto:2013kqa}
  N.~Armesto, J.~Rojo, C.~A.~Salgado and P.~Zurita,
  JHEP {\bf 1311} (2013) 015
  [arXiv:1309.5371 [hep-ph]].

\bibitem{Paukkunen:2014zia}
  H.~Paukkunen and P.~Zurita,
  JHEP {\bf 1412} (2014) 100
  [arXiv:1402.6623 [hep-ph]].

\bibitem{Kusina:2016fxy}
  A.~Kusina {\it et al.},
  Eur.\ Phys.\ J.\ C {\bf 77} (2017) no.7,  488
  [arXiv:1610.02925 [nucl-th]].
  
\bibitem{Chekanov:2007ac}
  S.~Chekanov {\it et al.} [ZEUS Collaboration],
  Phys.\ Rev.\ D {\bf 76} (2007) 072011
  [arXiv:0706.3809 [hep-ex]].


\end{thebibliography}
\end{document}